\documentstyle[psfig,graphicx]{mn}
\begin{document}
\newcommand{\lsim}{\mathrel{\rlap{\raise -.3ex\hbox{${\scriptstyle\sim}$}}%
                   \raise .6ex\hbox{${\scriptstyle <}$}}}%
\newcommand{\gsim}{\mathrel{\rlap{\raise -.3ex\hbox{${\scriptstyle\sim}$}}%
                   \raise .6ex\hbox{${\scriptstyle >}$}}}%
\def\simlt{\mathrel{\rlap{\lower 3pt\hbox{$\sim$}}
        \raise 2.0pt\hbox{$<$}}}
\def\simgt{\mathrel{\rlap{\lower 3pt\hbox{$\sim$}}
        \raise 2.0pt\hbox{$>$}}}

\title[The large scale clustering of radio sources] {The large scale
  clustering of radio sources}

\author[M. Negrello, M. Magliocchetti, G. De Zotti]
{\parbox[t]{\textwidth} {M.~Negrello$^{1}$,
M.~Magliocchetti$^{2,1}$, G.~De Zotti$^{3,1}$}
\vspace*{6pt} \\
$~$ \\
$^1$SISSA, Via Beirut 4, I-34014, Trieste, Italy \\
$^2$INAF -- Osservatorio Astronomico di Trieste, Via Tiepolo 11, I-34131,
Italy \\
$^3$INAF -- Osservatorio Astronomico
di Padova, Vicolo dell'Osservatorio 5, I-35122 Padova, Italy \\ }

\maketitle
\begin{abstract}

The observed two-point angular correlation function, $w(\theta)$,
of mJy radio sources exhibits the puzzling feature of a power-law
behaviour up to very large ($\sim 10^\circ$) angular scales which
cannot be accounted for in the standard hierarchical clustering
scenario for any realistic redshift distribution of such sources.
After having discarded the possibility that the signal can be
explained by a high density local ($z\simlt 0.1$) source
population, we find no alternatives to assuming that -- at
variance with all the other extragalactic populations studied so
far, and in particular with optically selected quasars --
radio sources responsible for the large-scale clustering signal
were increasingly {\it less} clustered with increasing look-back
time, up to at least $z \simeq 1$. The data are accurately
accounted for in terms of a bias function which decreases with
increasing redshift, mirroring the evolution with cosmic time of
the characteristic halo mass, $M_\star$, entering the non linear
regime. In the framework of the `concordance cosmology', the
effective halo mass controlling the bias parameter is found to
decrease from about $10^{15}\,M_\odot/h$ at $z\simeq 0$ to the
value appropriate for optically selected quasars, $\simeq
10^{13}\,M_{\odot}/h$, at $z\simeq 1.5$. This suggests that, in
the redshift range probed by the data, the clustering evolution of
radio sources is ruled by the growth of large-scale structure, and
that they are associated with the densest environments virializing
at any cosmic epoch. The data provide only loose constraints on
radio source clustering at $z \simgt 1$ so we cannot rule out
the possibility that at these redshifts the clustering evolution
of radio sources enters a different regime, perhaps similar to
that found for optically selected quasars. The dependence of the
large-scale shape of $w(\theta)$ on cosmological parameters is
also discussed.

\end{abstract}

%
%%%%%%%%%%%%%%%%%%%%%%%%%%%%%%%%%%%%%%%%%%%%%%%%%%%%%%%%%%%%%%%%%%%%%%%%%
%
\section{Introduction}

Extragalactic radio sources are well suited to probe the large
scale structure of the Universe since they are detected over
large cosmological distances (up to $z\sim 6$), are unaffected by
dust extinction, and can thus provide an unbiased sampling of
volumes larger than those usually probed by optical surveys. On
the other hand, their 3D-space distribution can be recovered only
in the very local Universe ($z\lsim0.1$; see Peacock $\&$
Nicholson 1991; Magliocchetti et al. 2004) because the majority of
radio galaxies detected in the available large area surveys,
carried out at low frequencies, have very faint optical
counterparts, so that measurements of their redshifts are a
difficult task. As a result, only the angular clustering can be
measured for the radio source population. Interestingly,
high-frequency surveys have much higher identification rates
(Ricci et al. 2004), suggesting that this difficulty may be
overcome when such surveys will cover sufficiently large areas.

Even the detection of clustering in the 2D distribution of radio
sources proved to be extremely difficult (see Webster 1976,
Seldner $\&$ Peebles 1981, Shaver $\&$ Pierre 1989) since, when
projected onto the sky, the space correlation is significantly
diluted because of the broad redshift distribution of radio
sources. It was only with the advent of deep radio surveys
covering large areas of the sky, such as the FIRST (Becker, White
$\&$ Helfand 1995), WENSS (Rengelink et al. 1997), NVSS (Condon et
al. 1998), and SUMSS (Bock, Large $\&$ Sadler 1999), that
the angular clustering of this class of objects has been detected
with a high statistical significance down to flux density limits
of few mJy (see Cress et al. 1997 and Magliocchetti et al. 1998,
1999 for the FIRST survey; Blake $\&$ Wall 2002a,b and Overzier et
al. 2003 for NVSS; Rengelink $\&$ R{\"o}ttgering 1999 for
WENSS and Blake, Mauch $\&$ Sadler 2004 for SUMSS). Amongst
all the above surveys, NVSS is characterized by the most extensive
sky coverage and can thus take advantage of high statistics
despite its relatively high flux limit ($\sim 3\,$mJy vs $\sim
1\,$mJy of FIRST). The two-point angular correlation function,
$w(\theta)$, measured for NVSS sources brighter than 10 mJy is
well described by a power-law, extending from $\sim 0.1$ degrees
up to scales of almost 10 degrees. A signal of comparable
amplitude and shape was detected in the FIRST survey at the same
flux density limit, on scales of up to 2-3 degrees (see e.g.
Magliocchetti et al. 1999), while at larger angular separations
any positive clustering signal - if present - was hidden by the
noise.

Most analyses of radio source clustering performed so far (see
e.g. Blake $\&$ Wall 2002a,b; Overzier et al. 2003) assume a
two-point spatial correlation function of the form $\xi_{\rm
rg}(r)=(r/r_{0})^{-\gamma}$. The power-law shape is in fact
preserved when projected onto the sky (see Limber 1953), so that
the observed behaviour of the angular correlation is well
recovered. The studies summarized by Overzier et al. (2003)
typically found correlation lengths $r_{0}$ in the range
5--15$\,h^{-1}\,$Mpc. This large range may reflect on one hand
real differences in the correlation properties of radio sources of
different classes/luminosities, and, on the other hand, the large
uncertainties on both the redshift distribution of the sources and
the time-evolution of clustering, which are necessary ingredients
to estimate $\xi_{\rm rg}(r)$ from the observed $w(\theta)$.

Overzier et al. (2003) found that the observed clustering of
powerful radio sources is consistent either with an essentially
redshift independent {\it comoving} correlation length $r_0=14 \pm
3\,h^{-1}\,$Mpc, close to that measured for extremely red objects
(EROs) at $z\simeq 1$ (Daddi et al. 2001) {\it or} with a galaxy
conservation model, whereby the clustering evolution follows the
cosmological growth of density perturbations. In the latter case,
the present-day value of $r_0$ for the most powerful radio sources
is comparable to that of local rich clusters. 

A deeper examination of the power-law behaviour of the angular
two-point correlation function up to scales of the order of $\sim
10^\circ$ highlights some interesting issues.  In fact, within the
Cold Dark Matter paradigm of structure formation, the spatial
correlation function of matter displays a sharp cut-off around a
comoving radius of $r \sim 100\,\hbox{Mpc}$\footnote{We assume a
flat universe with a cosmological constant and
$\Omega_{0,m}=0.27$, $\Omega_{b}=0.045$, $\sigma_{8}=0.9$, $n=1$,
and $h=0.72$, in agreement with the first-year WMAP results
(Spergel et al. 2003).} (see e.g. Matsubara 2004, Fig.~1),
which, at the average redshift for radio sources $<z>\sim 1$,
corresponds to angular separations of only a few ($\sim
1^\circ-2^\circ$) degrees, in clear contrast with the
observations. This opens the question of how to reconcile the
clustering properties of these sources with the standard scenario
of structure formation. Magliocchetti et al. (1999) claim that the
large-scale positive tail of the angular correlation function
$w(\theta)$ can be reproduced by allowing for a suitable choice of
the time-evolution of the bias parameter, characterizing the way
radio galaxies trace the underlying mass distribution. Although
promising, this approach suffers from a number of limitations due
to both theoretical modelling and the quality of data then
available.

The aim of the present work is therefore to investigate in better
detail and to provide a self-consistent explanation of the puzzling
behaviour of the angular correlation function of radio sources,
especially on large angular scales. We will concentrate on the results
from the NVSS survey (Blake \& Wall 2002b; Overzier et al. 2003) as,
thanks to the extremely good statistics, a clear detection of a
positive clustering signal was obtained up to $\sim 10^\circ$.  We
will exploit the available spectroscopic information on local radio
sources to limit the uncertainties on their redshift distribution, and
will mainly focus on the time evolution of their clustering properties
via the bias parameter. In this way we will derive interesting
constraints on the typical mass of dark matter halos hosting the
population of radio sources. We will also investigate the dependence
of the predicted angular correlation function on cosmological
parameters.

The layout of the paper is as follows. A short description of the NVSS
survey is presented in Section~\ref{sec:NVSS}.  Section~\ref{sec:Nz}
illustrates the adopted model for the redshift distribution of mJy
radio sources, while in Section~\ref{sec:model} we provide the
formalism for the two-point angular correlation function. Results and
discussions are presented in Section~\ref{sec:results}. In
Section~\ref{sec:conclusions} we summarize our main conclusions.

%
%%%%%%%%%%%%%%%%%%%%%%%%%%%%%%%%%%%%%%%%%%%%%%%%%%%%%%%%%%%%%%%%%%%%%
%
\section{The NVSS survey}
\label{sec:NVSS}
The NVSS (NRAO VLA Sky Survey, Condon et al. 1998) is the largest
radio survey that currently exists at $1.4\,$GHz. It was
constructed between 1993 and 1998 and covers $\sim10.3\,$sr of the
sky north of $\delta=-40^{\circ}$. The survey was performed with
the VLA in the D configuration and the full width at half-maximum
(FWHM) of the synthesized beam is 45 arcsec. The source catalogue
contains 1.8$\times$10$^6$ sources and it is claimed to be 99 per
cent complete above the integrated flux density S$_{\rm
1.4GHz}=3.5\,$mJy.

The two-point angular correlation function, $w(\theta)$, of NVSS
sources has been measured by Blake $\&$ Wall (2002a, 2002b) and
Overzier et al. (2003) for different flux-density thresholds
between 3 mJy and 500 mJy. The overall shape of $w(\theta)$ is
well reproduced by a double power-law. On scales below $\sim
0.1^\circ$, the steeper power-law reflects the distribution of the
resolved components of single giant radio sources. On larger
scales the shallower power-law describes the correlation between
distinct radio sources. Since the behaviour on small scales is
mainly determined by the joint effect of the astrophysical
properties of the sources and of the resolution of VLA in the
various configurations, while that on larger scales is of
cosmological origin, we will concentrate on the latter only. In
fact, the clustering behaviour on scales $\simgt 0.1^\circ$
provides insights on both the nature of the radio sources, through
the way in which they trace the underlying dark matter
distribution, and on the cosmological framework which determines
the distribution of dark matter at each epoch.

We will consider the two-point angular correlation function as
measured by Blake $\&$ Wall (2002b) for sources with $S_{1.4GHz}\geq
10$ mJy. Such a flux limit ensures the survey to be complete and, at
the same time, provides good enough statistics to measure angular
clustering up to very large scales with small uncertainties. Moreover,
this limit also represents the deepest flux-density for which
systematic surface density gradients are approximately negligible
(Blake \& Wall 2002a). The NVSS source surface density at this
threshold is 16.9 deg$^{-2}$.

The data corresponding to separations in the range
$0.1^{\circ}\simlt\theta\simlt0.3^{\circ}$ most likely suffer from
a deficit of pairs probably due to an imperfect cleaning of bright
side lobes (see Blake, Mauch $\&$ Sadler, 2004). Therefore in the
following we will only consider scales $\theta > 0.3^\circ$.
In this range of scales, the measured angular correlation function
is found to be described as a power law, $w(\theta)=1.49\cdot
10^{-3}\times \theta^{-1.05}$, with $\theta$ in degrees (see
Blake, Mauch $\&$ Sadler 2004).

%
%
%%%%%%%%%%%%%%%%%%%%%%%%%%%%%%%%%%%%%%%%%%%%%%%%%%%%%%%%%%%%%%%%%%%%%%%
%
\section{Redshift distribution of milliJansky radio sources}
\label{sec:Nz}

\begin{figure}
\centering \vspace{7.cm} \includegraphics{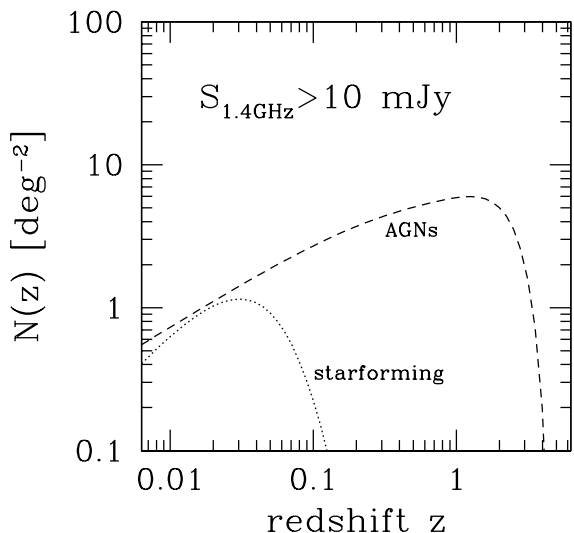} \caption{Adopted redshift
distribution per unit redshift interval, ${\mathcal N}(z)$, for the
radio source population with $S_{1.4{\rm GHz}}\ge 10\,$mJy. The dashed
line represents the contribution of AGN-powered radio sources
according to the PLE model of DP90, while the dotted line shows the
contribution from star-forming galaxies obtained by Magliocchetti et
al. (2002).}
\label{fig:dNdz_1.4GHz_10mJy}
\end{figure}

In order to provide theoretical predictions for the angular
two-point correlation function of a given class of objects it is
necessary to know their redshift distribution, ${\mathcal N}(z)$,
i.e. the number of objects per unit comoving volume as a function
of redshift. Unfortunately, the redshift distribution of mJy radio
sources is not yet accurately known as the majority of radio
sources powered by Active Galactic Nuclei (AGN) -- which dominate
the mJy population -- are located at cosmological distances
($<z>\sim1$) and are in general hosted by galaxies which are
optically extremely faint.

A large set of models for the epoch-dependent Radio Luminosity
Function (hereafter RLF) are available in literature (see e.g.  Dunlop
$\&$ Peacock 1990, Rowan-Robinson et al. 1993, Toffolatti et al. 1998,
Jackson $\&$ Wall 1999, Willot et al. 2001), but they all suffer from
the fact that they are mainly based on, and constrained by data sets
which include only bright sources ($S_{1.4GHz}\gsim 100\,$mJy) so that
any extrapolation of their predictions to lower fluxes is quite
uncertain. Amongst all the available RLFs, those provided by Dunlop
$\&$ Peacock (1990, hereafter DP90) are the most commonly used to
infer the redshift distribution of radio sources at the mJy
level. DP90 derived their set of RLFs on the basis of
spectroscopically complete samples from several radio surveys at
different frequencies. By using a `maximum entropy' analysis they
determined polynomial approximations to the luminosity function and
its evolution with redshift which were all consistent with the data
available at that time. In addition, they also proposed two models of
a more physical nature, assuming either Pure Luminosity Evolution
(hereafter PLE) or Luminosity/Density Evolution.

Recently, with the aim of determining the photometric and
spectroscopic properties for at least the population of local
(i.e. $z<0.2$) radio sources at the mJy level, a number of studies
have concentrated on samples obtained combining radio catalogues
like FIRST and NVSS with optical ones like SDSS and 2dFGRS (Sadler
et al. 2002, Ivezi$\acute{c}$ et al. 2002, Magliocchetti et al.
2002, 2004). The radio/optical samples obtained in this way
provide a crucial constraint at $z\sim 0$ for any theoretical
model aiming at describing the epoch-dependent RLF.

For instance, Magliocchetti et al. (2002, hereafter M02) have shown
that the RLF at 1.4 GHz derived from all the objects in their
spectroscopic sample having $S\geq1$ mJy and $b_{J}\leq19.45$, is well
reproduced at relatively high radio luminosities ($P_{1.4{\rm
GHz}}>10^{20.5}\,\hbox{W}\, \hbox{Hz}^{-1} \,\hbox{sr}^{-1}$) by the
DP90's PLE model for steep-spectrum FRI-FRII sources (Fanaroff
$\&$ Riley 1974). At lower radio luminosities, where the radio
population is dominated by star-forming galaxies, the measured RLF is
better described by the one proposed by Saunders et al. (1990) for
IRAS galaxies (see also Rowan-Robinson et al. 1993), although with a
small adjustment of the parameters.

Following to M02, we adopt the DP90 PLE model to derive the
redshift distribution of AGN-fuelled radio sources with
$S_{1.4{\rm GHz}}\ge 10\,$mJy (see
Fig.~\ref{fig:dNdz_1.4GHz_10mJy}, dashed line). We will not take
into account the other DP90 models, since we have found them to be
inconsistent with the local RLF. In fact, while the model assuming
density/luminosity evolution over-predicts the number of
steep-spectrum radio galaxies below $P_{1.4{\rm GHz}}\lsim 10^{22}
\,\hbox{W}\, \hbox{Hz}^{-1} \,\hbox{sr}^{-1}$, models using
polynomial approximations for the RLF give an unrealistic
overestimate of the number of local sources at every luminosity.

We will also use the fit provided by M02 to the local RLF in order
to estimate the redshift distribution of star-forming radio
galaxies with $S_{1.4{\rm GHz}}\ge10\,$mJy (see
Fig.~\ref{fig:dNdz_1.4GHz_10mJy}, dotted line).  Note that
star-forming galaxies comprise less than 30\% of the $z\lsim 0.1$
population of radio sources with $S_{1.4{\rm GHz}}\ge 10\,$mJy,
and only $\simeq 0.5\%$ of the total counts at this flux limit.
As we will show in the next Section, this result implies that the
contribution of star-forming galaxies to the large scale angular
clustering of NVSS sources is totally negligible.

%
%%%%%%%%%%%%%%%%%%%%%%%%%%%%%%%%%%%%%%%%%%%%%%%%%%%%%%%%%%%%%%%%%%%%%%%
%
\section{model for the angular correlation function}
\label{sec:model}

The two-point angular correlation function, $w(\theta)$, of a
population of extragalactic sources is related to their spatial
correlation function, $\xi(r,z)$, and to their redshift
distribution via the Limber's (1953) equation:
\begin{eqnarray}
w(\theta)&=&\int_{{\mathcal Z}}dz{\mathcal N}^{2}(z)\int_{{\mathcal Z}^{\prime}(z)} d(\delta z)
\xi[r(\delta z,\theta),z] \nonumber \\
&~&\times \left[\int_{{\mathcal Z}}dz{\mathcal N}(z)\right]^{-2}.
\label{eq:wth_obj}
\end{eqnarray}
In this expression, $r(\delta z,\theta)$ is the comoving spatial
distance between two objects located at redshifts $z$ and
$z+\delta z$ and separated by an angle $\theta$ on the sky. For a
flat universe and in the small angle approximation (which is still
reasonably accurate for the scales of interest here, i.e.
$0.3^{\circ}\lsim\theta\lsim10^{\circ}$),
\begin{eqnarray}
r^{2}=\left(\frac{c}{H(z)}\right)^{2}\delta z^2+d^{2}_{\theta}(z),
\label{eq:r}
\end{eqnarray}
where $H(z)=H_{0}E(z)$ is the time-dependent Hubble parameter and
$d_{\theta}(z)$ is the comoving linear distance on the sky surface
corresponding, at a the redshift $z$, to an angular separation
$\theta$. Integrations in Eq.~(\ref{eq:wth_obj}) are
performed in the ranges ${\mathcal Z}=[z_{\rm min},z_{\rm max}]$
and ${\mathcal Z}^{\prime}(z)=[z_{\rm min}-z,z_{\rm max}-z]$,
where $z_{\rm min}=0$ and $z_{\rm max}=6$ are, respectively, the
minimum and the maximum redshift at which radio sources are
observed.

On sufficiently large scales (e.g. $r\gsim3\,$Mpc), where the
clustering signal is produced by galaxies residing in distinct
dark matter halos and under the assumption of a one-to-one
correspondence between sources and their host halos, the spatial
two-point correlation function can be written as the product of
the correlation function of dark matter, $\xi_{\rm DM}$, times the
square of the bias parameter, $b$ (Matarrese et al. 1997,
Moscardini et al. 1998):
\begin{eqnarray}
\xi(r,z)=b^{2}(M_{\rm eff},z)\xi_{\rm DM}(r,z).
\label{eq:xi_obj}
\end{eqnarray}
Here, $M_{\rm eff}$ represents the effective mass of the dark matter
haloes in which the sources reside and $b$ is derived in the extended Press
\& Schechter (1974) formalism according to the prescriptions of
Sheth \& Tormen (1999).

The function $\xi_{\rm DM}$ is determined by the power spectrum of
primordial density perturbations as well as by the underlying
cosmology. For the power spectrum of the primordial fluctuations
we adopt the fitting relations by Eisenstein $\&$ Hu (1998) which
account for the effects of baryons on the matter transfer
function. The initial power spectrum is assumed to be scale
invariant with a slope $n=1$. As already mentioned, we adopt a
`concordance cosmology', in agreement with the first-year WMAP
results (Spergel et al. 2003).

In the range of scales of interest here the clustering evolves in
the linear regime, so that $\xi_{\rm
DM}(r,z)=D^{2}(z)\xi_{DM}(r,0)$, $D(z)$ being the linear density
growth-rate (Carroll, Press $\&$ Turner 1992).

Since the population of radio sources with $S_{\rm
1.4GHz}\geq10$ mJy is composed by two different types of objects,
i.e. AGNs and star-forming galaxies, which display different
clustering properties (see e.g. Saunders, Rowan-Robinson $\&$ Lawrence
1992, Madgwick et al. 2003), the angular correlation function of
the whole NVSS sample is given by [cf. e.g. Wilman et al. 2003,
Eq.~(9)]:
\begin{equation}
w(\theta)=f^{2}_{\rm AGN}w_{\rm AGN}(\theta)+f^{2}_{\rm SF}w_{\rm
SF}(\theta)+2f_{\rm AGN}f_{\rm SF}w_{\rm cross}(\theta)
\label{eq:wth_tot}
\end{equation}
where $f_{\rm AGN}$ and $f_{\rm SF}$ are the fractions of AGNs and
star-forming galaxies in the whole sample, respectively, i.e.
\begin{equation}
f_{\rm AGN/SF}=\frac{\int_{\mathcal Z}dz{\mathcal N}_{\rm
AGN/SF}(z)}{\int_{\mathcal Z}dz{\mathcal N}(z)}\ ,
\end{equation}
${\mathcal N}(z)$ being the global redshift distribution.

In Eq.~(\ref{eq:wth_tot}), $w_{\rm AGN}$ and $w_{\rm SF}$ are the
angular correlation functions of the two classes of radio sources,
while $w_{\rm cross}$ accounts for the cross-correlation between
the two populations:
\begin{eqnarray}
w_{\rm cross}(\theta) &=& \int_{\mathcal Z}dz{\mathcal N}_{\rm
    AGN}(z){\mathcal N}_{\rm SF}(z) \nonumber \\
&~& \times \int_{{\mathcal Z}^{\prime}(z)} d(\delta z)\xi_{{\rm cross}}[r(\delta z,\theta),z] \nonumber \\
&~& \times \left[ \int_{\mathcal Z}dz{\mathcal N}_{\rm
    AGN}(z)\int_{\mathcal Z}dz{\mathcal N}_{\rm SF}(z)
    \right]^{-1}.
\label{eq:wth_cross}
\end{eqnarray}
We model $\xi_{\rm cross}$ as [cf. e.g. Magliocchetti et al. 1999,
Eq.~(31)]:
\begin{eqnarray}
\xi_{\rm cross}(r,z) &=& \sqrt{\xi_{\rm AGN}(r,z)\xi_{\rm SF}(r,z)}
\nonumber \\
&=& b(M^{\rm AGN}_{\rm eff},z)b(M^{\rm SF}_{\rm eff},z)\xi_{\rm
  DM}(r,z);
\label{eq:xi_cross}
\end{eqnarray}
where $\xi_{\rm AGN}$ and $\xi_{\rm SF}$ are the spatial
correlation functions of AGNs and star-forming galaxies respectively, while
$M^{\rm AGN}_{\rm eff}$ and $M^{\rm SF}_{\rm  eff}$ denote the
effective masses of the corresponding dark matter halos [cf.
Eq.~(\ref{eq:xi_obj})]. Note that this definition likely
overestimates the cross-correlation term, which may be close to
zero, since AGN-powered radio sources are normally found in
clusters while star-forming galaxies preferentially reside in the field.

%
%%%%%%%%%%%%%%%%%%%%%%%%%%%%%%%%%%%%%%%%%%%%%%%%%%%%%%%%%%%%%%%%%
%

\section{Results}
\label{sec:results}

\begin{figure}
\centering \vspace{9.0cm} \includegraphics{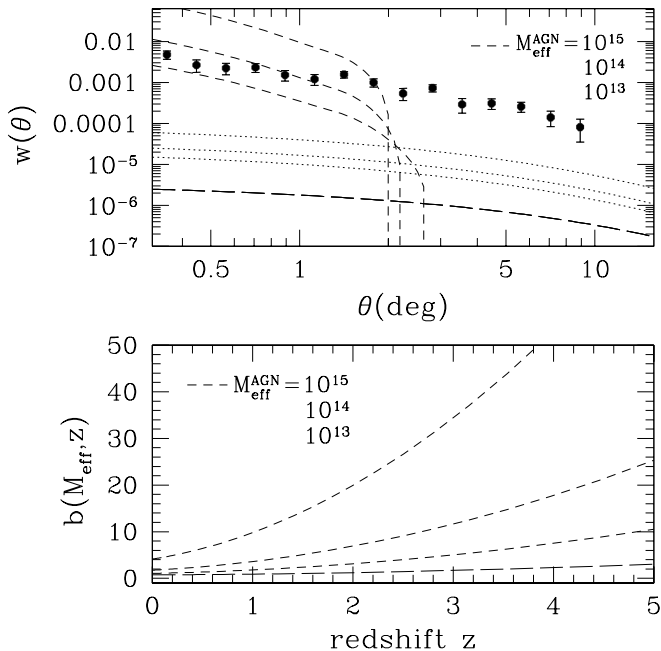} \caption{ {\it
Top panel}: contributions to the two-point angular correlation
function of NVSS sources with $S_{\rm 1.4GHz}\geq10$ mJy from AGNs
(short-dashed curves), star-forming galaxies (long-dashed curve),
and their cross-correlation (dotted curves). The effective halo
mass of AGN-powered radio galaxies is assumed to be constant
with redshift. The three
short-dashed and dotted curves correspond, from bottom to
top, to M$^{\rm AGN}_{\rm eff} = 10^{13}$, $10^{14}$ and
$10^{15}\,$M$_{\odot}$/h, respectively. The effective mass of the
halo in which star-forming galaxies reside is M$^{\rm SF}_{\rm
eff} = 10^{11}$ M$_{\odot}$/h (long-dashed curve). Points with
error bars represent the angular two-point correlation function of
NVSS sources as measured by Blake $\&$ Wall (2002b). {\it Lower
panel}: Evolution of the bias parameter for AGNs (short-dashed
curves) and star-forming galaxies (long-dashed curves).}
\label{fig:wth_Meffconst}
\end{figure}

As a first step, we obtain estimates of the angular two-point 
correlation function, $w(\theta)$, assuming a redshift-independent
effective mass for the host halos of AGN-powered radio
sources, $M^{\rm AGN}_{\rm eff}$, consistent with the clustering
properties of optically selected quasars (Porciani, Magliocchetti
\& Norberg 2004; Croom et al. 2005). The value of $M^{\rm
AGN}_{\rm eff}$ is taken as a free parameter. The observationally
determined spatial correlation length of star-forming galaxies,
$r_{0}(z=0)\sim 3$--4 Mpc/h (Saunders, Rowan-Robinson $\&$
Lawrence 1992; see also Wilman et al. 2003) is consistent with an
effective halo mass not exceeding $M^{\rm SF}_{\rm eff}=10^{11}$
$M_{\odot}$/h. We adopt this value in our analysis. Any dependence
of $M^{\rm SF}_{\rm eff}$ on redshift can be ignored because of
the small range in redshift covered by star-forming galaxies with
$S_{\rm 1.4GHz}\geq10$ mJy. 

In the top panel of Fig.~\ref{fig:wth_Meffconst} we show the
contributions to $w(\theta)$ of NVSS sources with $S_{1.4{\rm
GHz}}\ge 10\,$mJy arising from the
clustering of AGNs with $M^{\rm AGN}_{\rm eff} = 10^{13}$,
$10^{14}$, and $10^{15}\, M_{\odot}$/h (short-dashed curves) and
star-forming galaxies (long-dashed curve), as well as from the 
cross-correlation between these two populations (dotted curves). As
noted above, the cross correlation term, computed from
Eq.~(\ref{eq:xi_cross}) for the three values of $M^{\rm AGN}_{\rm
eff}$, is likely overestimated. The above contributions are then compared
with the observational determination of Blake $\&$ Wall (2002b).
In the lower panel of the same Figure the redshift-evolution of
the bias parameter is represented.

Figure~\ref{fig:wth_Meffconst} shows that the contribution of
star-forming galaxies to the observed angular correlation function is
negligible. This is because their contribution to the $S_{\rm
1.4GHz}\geq10$ mJy counts is small, implying $f_{\rm
SF}\sim5\times 10^{-3}$ [see Eq.~(\ref{eq:wth_tot})], compared
with $f_{\rm AGN}\sim 1$. We note that this conclusion is in disagreement
with that of Cress $\&$ Kamionkowsky (1998), which relies on the adoption of a
local RLF of star-forming galaxies noticeably exceeding the direct
estimates of M02.

\begin{figure}
\centering \vspace{4.8cm} \includegraphics{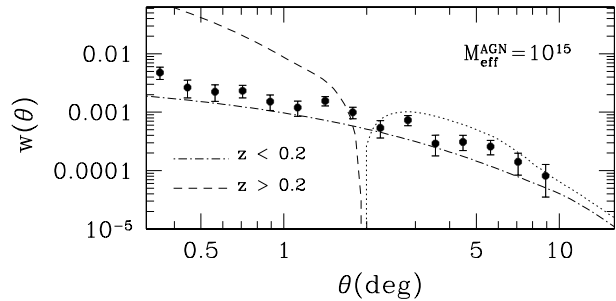}
\caption{Contributions to the two-point angular correlation function
of NVSS sources with $S_{\rm 1.4GHz}\geq10$ mJy from radio sources
above (dashed and dotted lines) and below (dot-dashed line) $z=0.2$,
for $M^{\rm AGN}_{\rm eff}=10^{15}\, M_{\odot}$/h and $M^{\rm SF}_{\rm
eff}=10^{11}\, M_{\odot}$/h. Star-forming galaxies contribute (yet
marginally) only at $z<0.2$. The contribution of higher redshift
sources (AGNs only) is positive for $\theta \simlt 2^\circ$, and becomes
negative on larger scales (the dotted line shows its absolute
values). } \label{fig:wth_z}
\end{figure}

Clearly, models in which the mass of the halo hosting AGNs is
constant in redshift badly fail at reproducing the overall shape
of the observed angular correlation function. This is because
contributions to $w_{\rm AGN}(\theta)$ on a given angular scale
come from both local and high-redshift sources. For a
redshift-independent halo mass, on scales $\theta\gsim 2^\circ$
the positive contribution of local sources is overcome by the
negative contribution of distant AGNs, which dominate the redshift
distribution and whose bias parameter rapidly increases with
redshift [see Eq.~(\ref{eq:xi_obj}) and
Fig.~\ref{fig:wth_Meffconst}]. In fact, for redshifts $z\gsim 0.5$--1
such angular scales correspond to distances where $\xi_{\rm
AGN}(r)$ is negative.

This situation is illustrated by Fig.~\ref{fig:wth_z}, showing
that the contribution to $w_{\rm AGN}(\theta)$ from local AGNs
($z< 0.2$, dot-dashed line) with $M^{\rm AGN}_{\rm eff}=10^{15}\,
M_{\odot}$/h comes close to accounting for the observed angular
correlation function (the dot-dashed line also includes the
contribution of star-forming galaxies with $M^{\rm SF}_{\rm
eff}=10^{11}\, M_{\odot}$/h, which however turns out to be almost
negligible). But if the same value of $M^{\rm AGN}_{\rm eff}$
also applies to high-redshift AGNs,  $w_{\rm AGN}(\theta)$ turns
out to be too high on small scales and negative on large scales.
Therefore, the only way to account for the behaviour of the data on
large scales is to weight down the high-$z$ contribution by 
decreasing the bias factor at such redshifts.
We note that a similar result -- which drastically rules out models 
featuring a bias function strongly increasing with z -- was obtained by 
Magliocchetti et al. (1999) via their counts in cells analysis of the FIRST data.

\begin{figure}
\includegraphics[width=1.045\columnwidth]{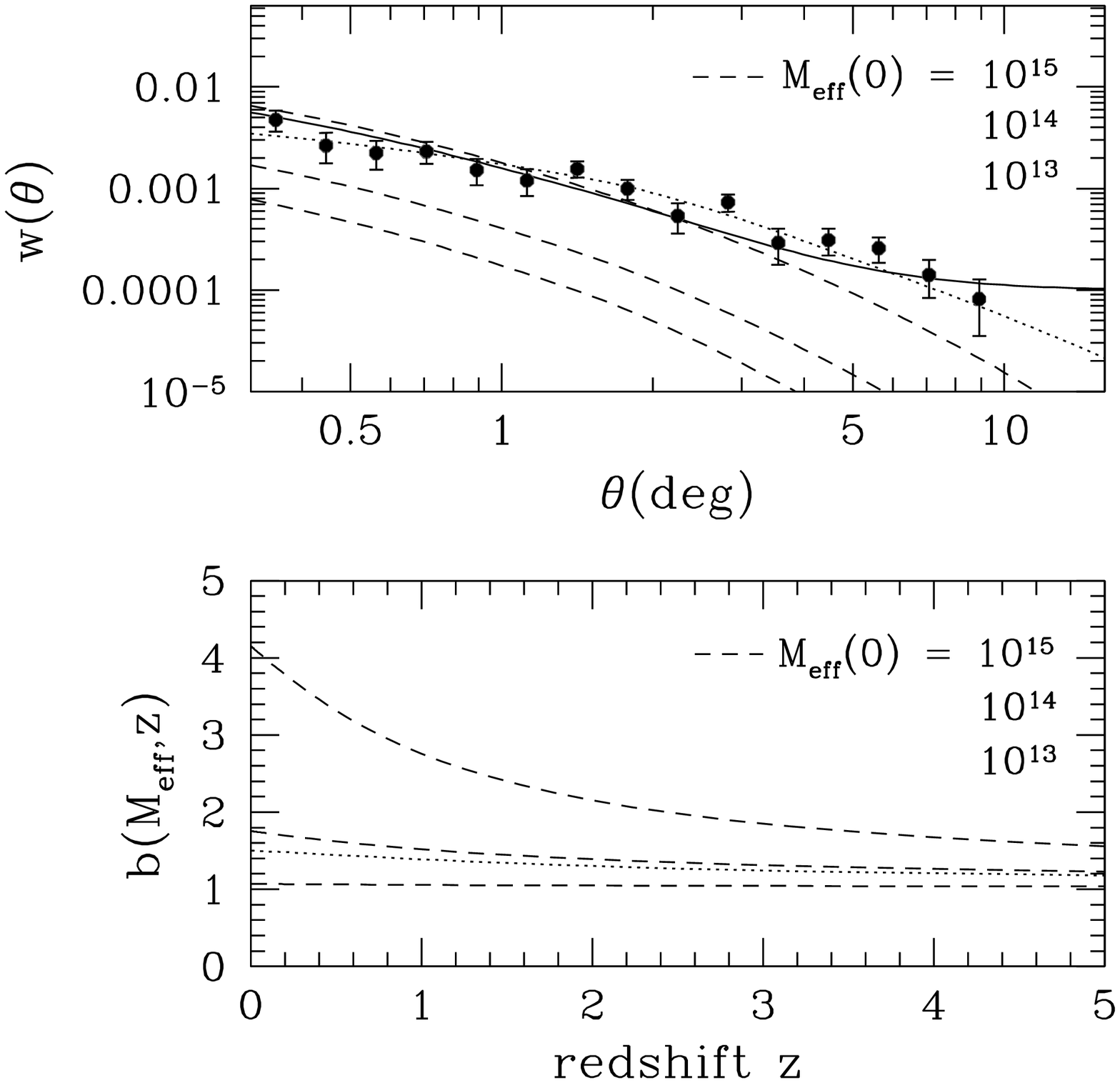}\hspace{0.2
cm} \includegraphics[width=1.045\columnwidth]{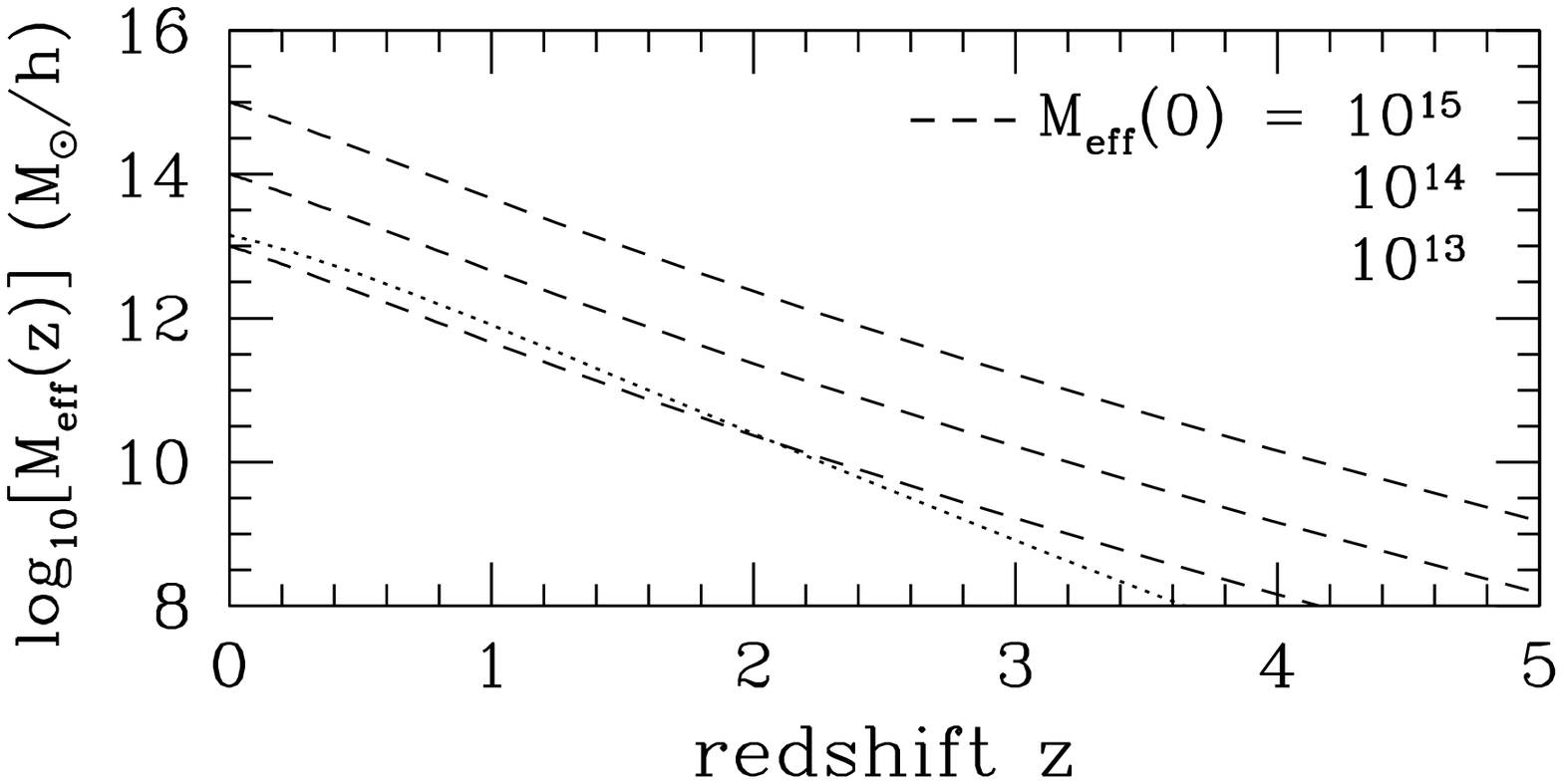}
\caption{Two-point angular correlation function (top panel), bias
parameter (central panel), and effective mass of the host dark matter
halo (bottom panel) for radio sources with $S_{1.4\rm GHz}\ge 10\,$mJy
for a model in which $M_{\rm eff}\propto M_{\star}(z)$ (see text for
details). The three dashed curves correspond, from bottom to top, to
different local values of the effective mass: M$_{\rm eff}(z=0)$ =
$10^{13}$, $10^{14}$ and $10^{15}\, M_{\odot}$/h. The solid line in
the top panel represents the prediction obtained for $M_{\rm eff}(z=0)
= 10^{14.9}\, M_{\odot}$/h by adding a constant offset
$\epsilon=0.0001$ to $w(\theta)$ in order to account for the effect of
possible spurious density gradients in the survey. The dotted curves
represent the prediction for $\Omega_{0,m}=0.105$ and $M_{\rm
eff}(z=0)=10^{13.15}\,M_{\odot}$/h.  Points with error bars show the
two-point angular correlation function of NVSS sources as measured by
Blake $\&$ Wall (2002b).}
\label{fig:wth_Meffz}
\end{figure}

Having shown that the contribution of star-forming galaxies to
the clustering signal is negligible, hereafter we will only consider 
AGNs and will drop the `AGN' index from the effective mass.

Based on the fact that AGN-powered radio galaxies are mainly found
in very dense environments such as groups or clusters of galaxies,
we can then assume that the epoch-dependent effective mass controlling
their clustering properties, $M_{\rm eff}$ is proportional to the
characteristic mass of virialized systems, $M_{\star}(z)$, which
 decreases with increasing redshift. $M_{\star}(z)$ is
defined by the condition (Mo $\&$ White 1996):
\begin{eqnarray}
\sigma[M_{\star}(z),z]=\delta_{c}(z) \label{eq:Mstar}
\end{eqnarray}
where $\sigma(M,z)$ is the rms fluctuation of the density field at
redshift $z$ on a mass scale $M$, while $\delta_{c}(z)$ is the
threshold density for collapse of a spherical perturbation at
redshift $z$. $M_{\star}$ then represents the typical mass scale
at which the matter density fluctuations collapse to form bound
structures. As required by the data, this choice for $M_{\rm eff}$ 
(whose redshift dependence is shown in the bottom panel of
Fig.~\ref{fig:wth_Meffz}) leads to an almost redshift-independent
bias factor (central panel of Fig.~\ref{fig:wth_Meffz}) except for
the highest value of $M_{\rm eff}(0)$ for which we have a moderate
decrease of $b$ with increasing $z$.

This model successfully reproduces the observed $w(\theta)$ up to
scales of at least $\simeq 4^\circ$ (top panel of
Fig.~\ref{fig:wth_Meffz}). The best fit is obtained for $M_{\rm
eff}(z=0) = 10^{14.96\pm0.04}\,M_{\odot}$/h. On the largest
angular scales ($\theta\gsim 5^\circ$), the model correlation
function still falls short of the data. However, small
systematic variations in the source surface density due to
calibration problems at low flux densities may spuriously enhance
the estimate of $w(\theta)$ (Blake \& Wall 2002a). An offset as
small as $\epsilon=10^{-4}$ would be enough to remove the
discrepancy between the model and the data (solid line in the top
panel of Fig.~\ref{fig:wth_Meffz}).

On the other hand, if the measured $w(\theta)$ is unaffected by
spurious density gradients in the survey, it may be possible to
reconcile the model with the observed signal by exploiting the
dependence of the model predictions on cosmological parameters,
and notably on the matter density parameter, on the baryonic
fraction and on the Hubble constant (see Matsubara 2004, Fig.~1).
Decreasing $\Omega_{0,m}$ shifts the cut-off of the dark matter
spatial correlation function $\xi_{\rm DM}$ to larger scales,
while a non-null value of $\Omega_{b}$ determines a peak in the
correlation function just below the cut-off scale. The amplitude
of this peak is an increasing function of $\Omega_{b}$.

A change in the value of $h$ has an effect which is qualitatively similar
to that induced by changing $\Omega_{0,m}$, but of a significantly
smaller amplitude; for this reason we have decided to keep this
parameter fixed to the reference value $h=0.72$. We have also
ignored the dependence of the clustering pattern on the rms mass
fluctuation, $\sigma_8$: in principle this quantity contributes to
the overall normalization of the spatial clustering, but we have
checked that in our case the effect of changing it is almost
negligible. This is because the decrease of $b$ with increasing
$\sigma_8$ is compensated by the corresponding increase of the
effective mass (assumed to be proportional to $M_{\star}(z)$).
Therefore we also keep $\sigma_8=0.9$.

\begin{figure*}
\centering \vspace{15.0cm} \includegraphics{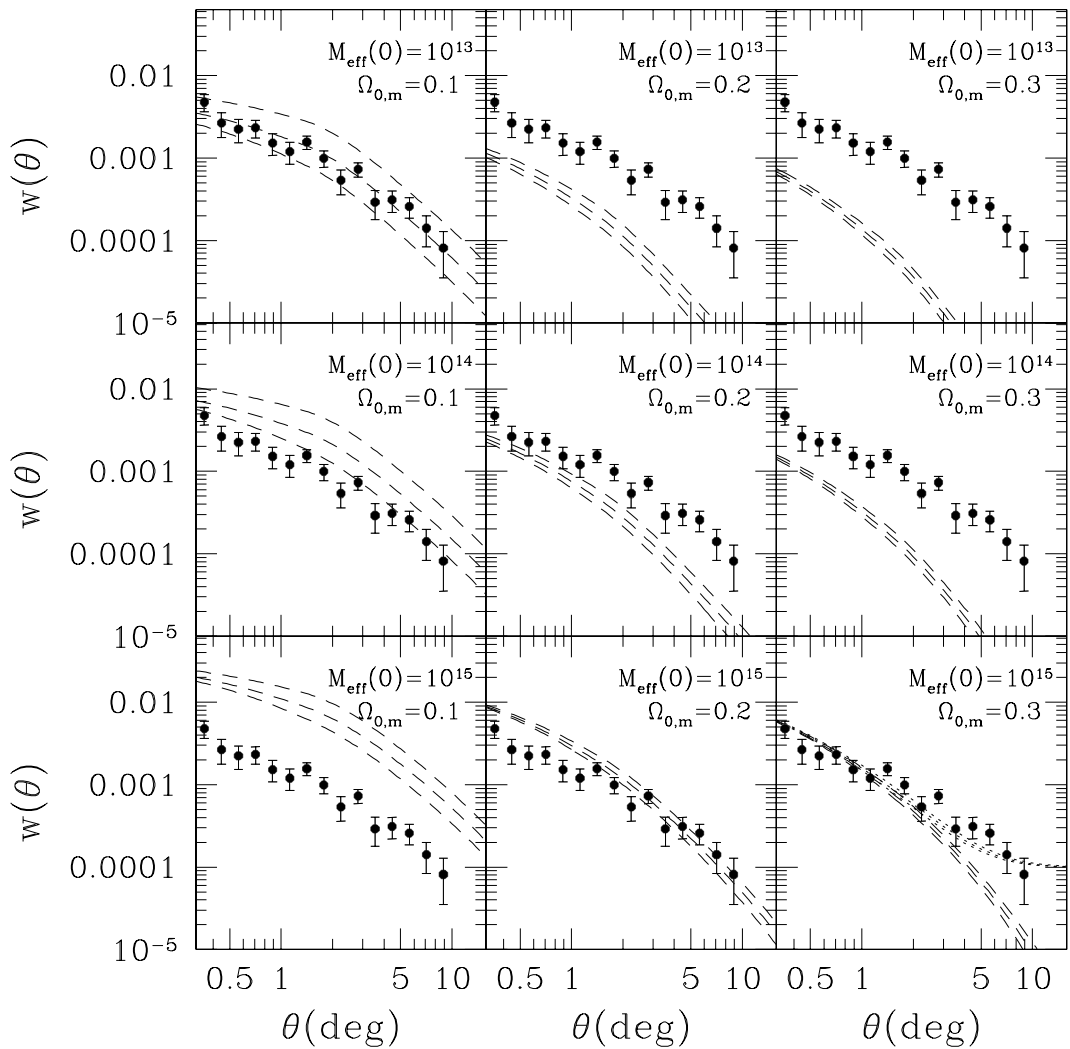} \caption{Dependence
on cosmological parameters of the two-point angular correlation
function of radio sources with $S_{1.4\rm GHz}\ge 10$ mJy. The
effective mass of the dark matter haloes in which the sources
reside is assumed to be proportional to $M_{\star}(z)$. In each
panel, the adopted values of $M_{\rm eff}(0)$ (in solar units) and
of the matter density parameter, $\Omega_{0,m}$ are indicated: the
former increases from top to bottom panels while the latter
increases from left-hand to right-hand panels. In each panel the
three dashed curves correspond to different choices for the
baryonic mass density parameter: $\Omega_{b}= 0.03$, 0.045 and
0.06 from the lower to the upper curves, respectively. In the
bottom right-hand panel the dotted curves (almost
indistinguishable from each other) represent the two-point
correlation function obtained by adding to the dashed curves a
constant offset $\epsilon=10^{-4}$ (see text for details).}
\label{fig:wth_table}
\end{figure*}

The parameters in our analysis are then: $M_{\rm eff}(0)$,
$\Omega_{0,m}$ and $\Omega_{b}$. In Fig.~\ref{fig:wth_table} we
show the predicted two-point angular correlation function for
different choices of these parameters. In each panel, the dashed
curves correspond (from bottom to top) to $\Omega_{b}=0.03$, 0.045
and 0.06. The values adopted for both $M_{\rm eff}(0)$, and
$\Omega_{0,m}$ are given in each panel: $M_{\rm eff}(0)$ increases
from the top to the bottom panels, while $\Omega_{0,m}$ increases
from the left-hand to the right-hand panels.

We note that variations of these parameters within the allowed
ranges mainly translate into changes of the normalization of
$w(\theta)$, although varying $\Omega_{b}$ also affects the $w(\theta)$ 
slope on scales
$\theta\simlt 2^\circ$, while changes in $\Omega_{0,m}$ modify 
its slope on larger scales.
In general, the amplitude of $w(\theta)$ increases with increasing
$M_{\rm eff}(0)$ or $\Omega_{b}$, and decreases for decreasing
$\Omega_{0,m}$. From Fig.~\ref{fig:wth_table} we can deduce that,
 given the observed $w(\theta)$, smaller values of $M_{\rm eff}(0)$ are
favoured for lower values of $\Omega_{0,m}$.

\begin{figure}
\centering \vspace{8.0cm} \includegraphics{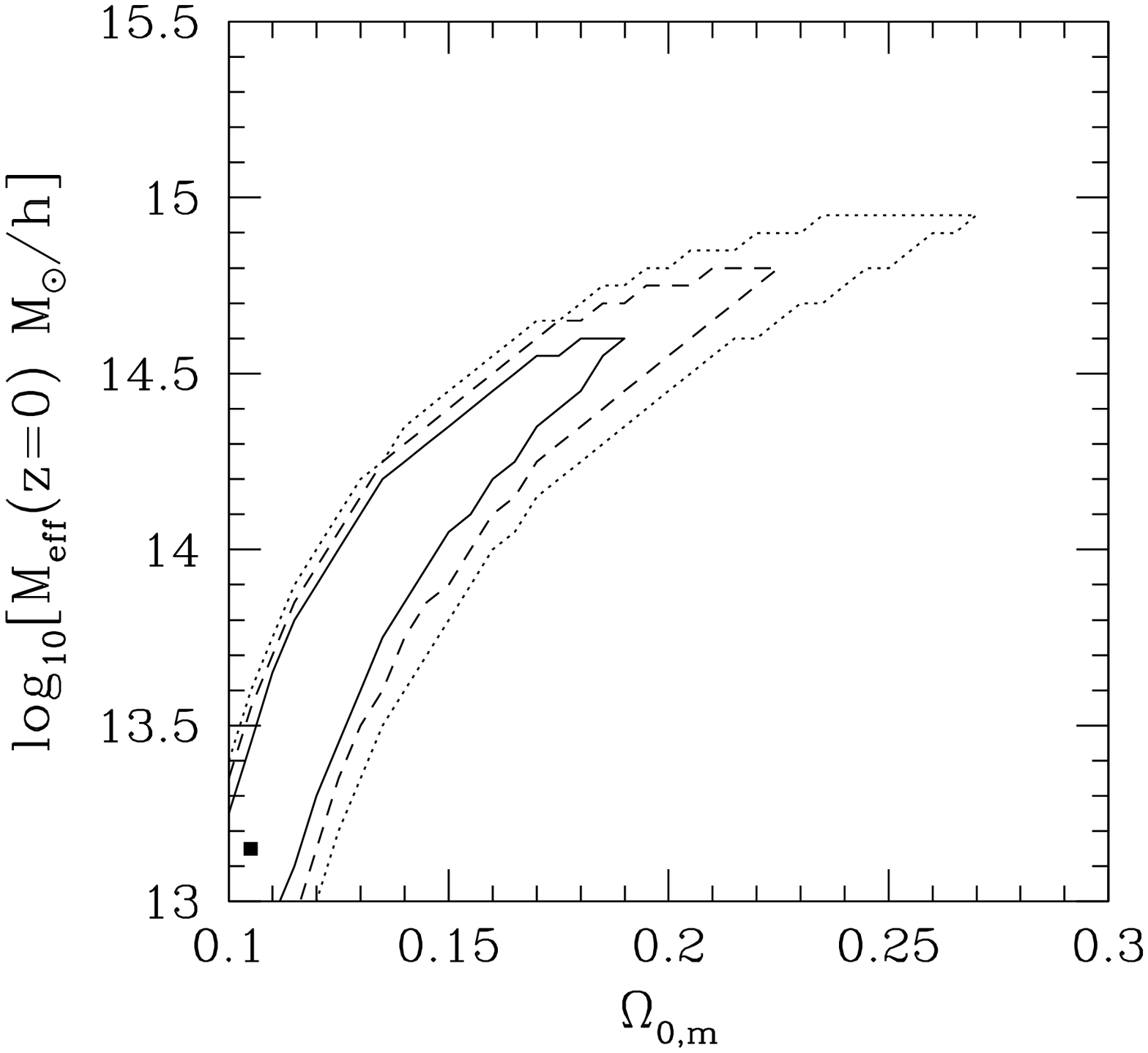} \caption{Confidence
contours derived from a $\chi^{2}$ analysis on the
$\Omega_{0,m}$-$\log_{10}[M_{\rm eff}(z=0)]$ plane, having assumed
$\Omega_b=0.045$. The curves, from inner to outer ones,
correspond to the 68.3, 95.4 and 99.7 per cent confidence
intervals.  The filled square corresponds to the best-fit set of
parameters $\Omega_{0,m}=0.105$, $M_{\rm eff}(z=0) = 10^{13.15}\,
M_{\odot}$/h.} \label{fig:grid_om0m.vs.Meffz}
\end{figure}

Fixing the baryonic content to the reference value
$\Omega_{b}=0.045$, we have then investigated the
$\Omega_{0,m}$--$M_{\rm eff}(0)$ interdependence by constructing
$\chi^2$ contours on the $\Omega_{0,m}$-$M_{\rm eff}(0)$ plane.
The results are shown in Fig.~\ref{fig:grid_om0m.vs.Meffz} where
the curves represent, from the innermost to the outermost,
contours corresponding to the 68.3, 95.4 and 99.7 per cent
confidence intervals, respectively. The filled square corresponds
to the best-fit set of parameters: $\Omega_{0,m}=0.105$, $M_{\rm
eff}(z=0) = 10^{13.15}\, M_{\odot}$/h. The resulting angular
two-point correlation function, and the redshift-evolution of both
the bias parameter and the effective mass are represented by the
dotted line in Fig.~\ref{fig:wth_Meffz}. We note that the effect
of a change in the baryonic content is just that of smearing the
relation between $\Omega_{0,m}$ and $M_{\rm eff}(0)$ (in
particular for $\Omega_{0,m}\lsim 0.2$) since there is a whole set
of $\Omega_{0,m}$--$M_{\rm eff}(0)$ pairs which can provide the same
best-fit to the data.

The fit obtained for a zero offset $\epsilon$ is really good, but
is obtained for a local matter density parameter $\Omega_{0,m}$
substantially smaller than what indicated by other data sets (see e.g.
Spergel et al. 2003), although the difference is significant to
less than $3\sigma$. Taken at face value, our results indicate that, within the
standard cosmological framework, in the local universe radio
sources are as strongly clustered as rich clusters of galaxies, in
excellent agreement with the findings of Peacock \& Nicholson
(1991) and Magliocchetti et al. (2004).

As a last comment, we note that the degeneracy between the host halo
mass and the matter density parameter indicates that large area
radio surveys without redshift information can hardly provide
significant constraints on cosmological parameters, unless
independent information on $M_{\rm eff}$ is available. On the
other hand, for a given set of cosmological parameters, such
surveys yield important information on the masses of the hosting halos.

\section{Discussion and Conclusions}
\label{sec:conclusions}

The observed angular two-point correlation function of mJy radio
sources exhibits the puzzling feature of a power-law behaviour up
to very large ($\sim 10^\circ$) angular scales. Standard models
for clustering, which successfully account for the clustering
properties of optically selected quasars, turn out to be unable to
explain the long positive tail of the $w(\theta)$, even when
`extreme' values for the parameters are invoked. This is because
-- according to the standard scenario for biased galaxy formation
in which extragalactic sources are more strongly clustered at
higher redshifts -- the clustering signal of radio sources at
$z\simgt 1$, which is negative on large scales, overwhelms that of
more local sources, yielding a sharp cut-off in the angular
two-point correlation function on scales of $\sim 1$--2 degrees.

The only way out we could find is to invoke the clustering strength of
radio sources to be weaker in the past. The data can then be accounted
for if we assume that the characteristic mass of the halos in
which these objects reside, $M_{\rm eff}(z)$, is proportional to
$M_{\star}(z)$, the typical mass-scale at which the matter density
fluctuations collapse to form bound structures (see Mo $\&$ White
1996).

A good fit of the observed $w(\theta)$ up to scales of at least
$4^\circ$ degrees is obtained for $M_{\rm eff}(z=0) =
10^{14.96\pm0.04}\,M_{\odot}$/h. The data on larger scales can be
accurately reproduced if the measured values of $w(\theta)$ are
slightly enhanced by small systematic variations in the source
surface density due to calibration problems at low flux densities
(Blake \& Wall 2002a). In the absence of such a systematic offset,
the data might indicate a lower value of the mean cosmic matter
density than indicated by other data sets. The best fit is
indeed obtained for $\Omega_{0,m}=0.105$ and $M_{\rm eff}(z=0) =
10^{13.15}\,M_\odot$/h, with rather large uncertainties on both
parameters, as shown by Fig.~\ref{fig:grid_om0m.vs.Meffz}. In
particular, the best fit value of $\Omega_{0,m}$ is less than
$3\sigma$ away from the `concordance' value. This shows that
current large scale radio surveys without redshift
measurements cannot provide strong constraints on cosmological
parameters, but are informative on the evolution history 
of the dark matter halos hosting radio sources.

\begin{figure}
\centering \vspace{8.0cm} \includegraphics{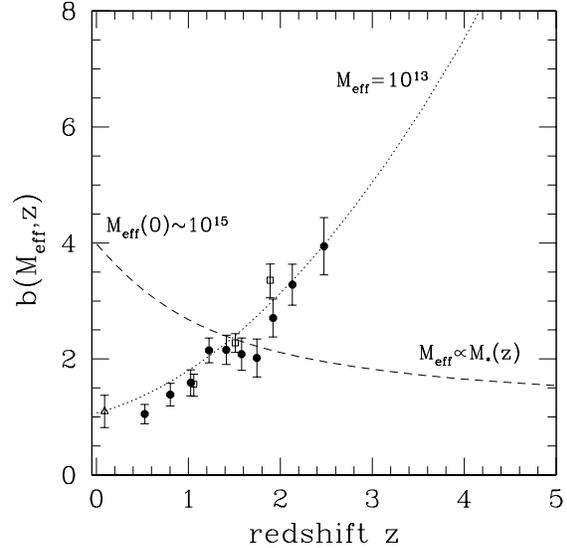} \caption{Bias factor as a function of
redshift. The data points are from observations of optically selected
quasars (open squares from Porciani et al.  2004, filled dots from
Croom et al. 2005, open triangle from Grazian et al. 2004; the values
of $b$ derived by these authors have been scaled to the values of
$\sigma_8$ and $\Omega_{0,m}$ used here), while the dashed line
represents the functional form derived in this work for the mJy
population of radio sources. The dotted curve shows the bias function
obtained for a characteristic halo mass constant in time and equal to
$\sim 10^{13}\,M_{\odot}$/h.} \label{fig:bz}
\end{figure}

Taken at face value, the above results point to different
evolutionary properties for different types of AGNs as the
decreasing trend for the effective mass ruling the clustering of
radio sources found in this work strongly differs from the
behaviour of that associated to optically selected quasars. 
For the latter sources,
the data are in fact consistent with a constant $M_{\rm eff}\sim
10^{13}\,M_{\odot}$/h, at least up to the highest probed redshift,
$z\simeq 2.5$  (see e.g. Grazian et al. 2004; Porciani et al. 2004;
Croom et al. 2005) This different behaviour is illustrated in
Fig.~\ref{fig:bz}, where the values of the bias inferred for optically
selected quasars have been scaled to the values of $\sigma_8$ and
$\Omega_{0,m}$ used here.

It must be noted, however, that the present data provide very weak
constraints on the clustering properties of radio sources at $z\simgt
1$. For example, we have directly checked that the predicted
$w(\theta)$ does not change significantly over the range of
angular scales considered here if we assume $M_{\rm eff}\propto
M_\star(z)$ for $z\lsim 1.5$ and $M_{\rm eff}\sim {\rm const}\sim
10^{13}\,M_{\odot}$/h at higher redshifts. It is thus possible
that the difference in the clustering evolution between
AGN-powered radio sources and optically selected quasars is
limited to $z\simlt 1$, consistent with observational
indications that, at higher redshifts, the environment of
radio-quiet and radio-loud QSOs is almost the same (Russel,
Ellison $\&$ Benn 2005).

Interestingly, our analysis indicates that, at $z\simeq 1$, the
effective halo mass associated to radio sources is consistent with
being essentially equal to that of optical quasars
(Croom et al. 2005). This suggests that, at least in this
redshift range, the bias parameter for radio-loud and radio-quiet AGNs is 
similar. 

On the other hand, our findings suggest that, at least for
$z\simlt 1$ and at variance with what found for optical quasars, the
clustering of radio sources reflects that of the largest halos
which collapse at any given cosmic epoch. This conclusion is
in keeping with results of previous studies showing that, locally,
radio sources are preferentially associated with groups and clusters of
galaxies (e.g. Peacock \& Nicholson 1991; Magliocchetti et al.
2004), and, at higher redshift, are often associated with very
massive galaxies and very massive galaxy environments (e.g.
Carilli et al. 1997; Best, Longair $\&$
R$\ddot{\mbox{o}}$ttgering 1998; Venemans et al. 2002; Best et al.
2003; Croft et al. 2005; Overzier et al. 2006).

The major intriguing point that remains unsolved is the
link with the population of optical QSOs. The clustering properties of
these latter objects seem to reflect that of ``normal'' elliptical galaxies. 
So why this difference? 
Clearly, more observations are crucial to shed light on this
issue. So far the main limitation to our understanding of the
environmental properties of radio sources has been due to
selection effects that allow identifications of mostly radio
galaxies in the local universe and exclusively quasars (regardless
of their radio activity) at higher redshifts. Therefore, it would
be of uttermost importance to consider a redshift range in which
the clustering properties of both radio galaxies and radio-active
quasars can be measured with good precision. We plan to tackle
this issue in a forthcoming paper.

\noindent
\section*{ACKNOWLEDGMENTS}
We warmly thank C. Blake and J. Wall for having provided, in a
tabular form, their estimates of the two-point angular correlation
function of NVSS sources and for clarifications on their analysis.
We acknowledge useful suggestions from the anonymous referee,
which helped to improve the paper. Work supported in part by MIUR
and ASI.

\end{document}